\documentclass[prl,amsmath,amssymb,showpacs,showkeys,twocolumn]{revtex4}
\usepackage{graphicx}
\usepackage{dcolumn}
\usepackage{rotating}
\usepackage{amssymb}
\usepackage{mathptmx}

\usepackage{amsfonts}
\usepackage{amsmath}\usepackage{bm}
\begin{document}

\title{ Majorana Fermions: Direct Observation in  $^3$He}

\author{Yu.~M.~Bunkov$^{(a)}$}
\author{R.R.~Gazizulin$^{(a,b)}$}

\affiliation{$^{(a)}$ CNRS, Inst NEEL, F-38042 Grenoble, France\\ Univ. Grenoble Alpes, Inst NEEL, F-38042 Grenoble, France\\
$^{(b)}$ Kazan  Federal University, Kazan, Russia}

\date{\today}

\keywords{Superfluid $^{3}$He, Heat capacity, Majorana fermions,
Universe in a Helium droplet.}

\maketitle

The Majorana fermion, a particle which act as its own antiparticles, was suggested by Majorana in 1937. A historic review of this problem can be found in \cite{Majorana}. The unambiguous observation of Majorana fermions remains an outstanding goal. No fundamental particles are known to be Majorana fermions, although there are speculations that the neutrino may be one. There are also theoretical suggestions that Majorana fermions may comprise a large fraction of cosmic Dark Matter \cite{DM}. It has also been predicted that certain topological media, which are gapped in the bulk, may have topologically protected surface excitations with the properties of Majorana fermions \cite{Wil}. This class of topological media includes topological insulators such as superfluid $^3$He-B, thin films of $^3$He-A, vortex lines in superfluids and superconductors \cite{Volovik14}, thin superconducting wires and the quantum vacuum of the Standard Model of the Universe \cite{VolovikUn}. Graphene may also belong to this class if a broad enough energy gap may appear due to spin-orbit interaction (or due to spontaneously broken symmetry) \cite{graphin}.  In this letter we report the  first direct observation of gapless Majorana quasiparticles which appear as Andreev bound states on the surface of superfluid $^3$He-B \cite{Volovik14} . We have made precise measurements of the heat capacity of superfluid $^3$He-B at the extreme low-temperatures limit. We are able to separate the heat capacity contributions of bulk Bogolyubov quasiparticles and the surface Majorana quasipartilces by their different temperature dependencies. We have found that at  0.11 mK the Majorana fermions contribute about  half of the bolometer heat capacity under the conditions of our experiments.

There are several scenarios for the experimental search for Majorana signatures \cite{1,11,2,2b,3}. Many investigations have focused on thin supeconducting wires. Here, the  region for Majorana formation is situated near the ends of wires, meaning that the Majoranas would have exactly zero energy and would be unable to move. The question is: how would it be possible to observe a not-moving zero-energy state, a difficult task.   There have been attempts to confirm the existence of such Majoranas indirectly by the observation of interference of Majorana states on each side of the wire through a supercondacting bridge  \cite{1}. However,  a better choice is offered by the topological insulator, the time-reversal invariant B-phase of superfluid $^3$He. The spin-triplet superfluid supports the existence of Majorana quasiparticles \cite{Volovik09,Volovik10}. This follows from the particle-hole symmetry of the Bogoliubov quasiparticles, $\gamma^+_E = \gamma_{-E}$.  Consequently at zero energy they satisfy the Majorana condition $\gamma^+_0 = \gamma_0 $. Zero-energy quasiparticle bound states appear in $^3$He-B when the underlying quasiparticle potential changes sign. This conditions is satisfied at edges, at surfaces and on vortices with odd integer winding numbers, where the quasiparticles form zero-energy Andreev bound states (a unique quasiparticle state in which Andreev reflection \cite{Volovik14,Andreev} plays a fundamental role).

The superfluid energy gap of $^3$He-B is suppressed near walls over a distance corresponding to the superfluid coherence length $\xi$, which is about 80 nm at zero bar. In contrast to superconducting wires, the Majorana quasiparticles in $^3$He-B are able to move along the surface of the sample and consequently have a kinetic energy. They have a linear dispersion relation with a zero gap, referred to as the Majorana cone \cite{2,2b}. Because of this specific dispersion relation, the Majorana quasiparticles have a non-zero energy and consequently have a heat capacity.  A proposed experimental method for observing the additional Majorana heat capacity can be found in \cite{suggest}, and the first attempt of its realization in \cite{Haplperin}. In the latter case the experiments were performed in superfluid $^3$He in a micron powder heat exchanger at a temperatures near $T_c$. Unfortunately, it is impossible to apply the Majorana theory quantitatively  to the results of these experiments.

The Majorana heat capacity obeys a power law with temperature, with a quadratic dependence on $T$ for the specular scattering of quasiparticles. The Bogolyubov quasiparticles in the bulk $^3$He have a gap and consequently the bulk heat capacity falls exponentially with reducing temperature. At some low temperature the heat capacity of the Bogolyubov quasiparticles becomes lower than that of the Majoranas one. This temperature depends on the ratio between the volume and the surface of the $^3$He sample in the bolometer. In the experiments described in this letter, we have observed a deviation from an exponential law for the quasiparticles heat capacity in bolometer, which corresponds well to an additional heat capacity due to zero-gapped Majorana quasiparticles. The very good agreement with theoretical predictions supports our claim of a direct observation of zero-gapped Majorana quasiparticles.

 \begin{figure}[htt]
 \includegraphics[width=0.4\textwidth]{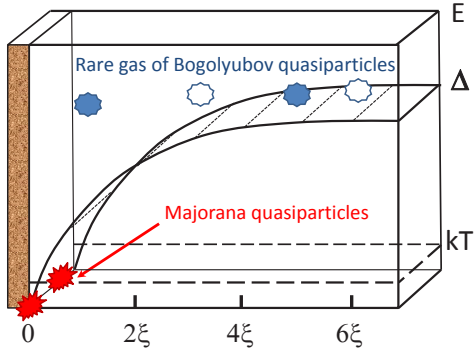}
 \caption{The experimental scenario. On the x axis we plot the distance from the $^3$He boundary in coherence length units.  The energy gap is shown by the solid line and temperature by the dashed line. Quasiparticles with energies above the gap can move in the bulk $^3$He. States with zero gap, the Majorana particles, only exist near the wall, but can move transversely along the walls with a kinetic energy which contributes an additional heat capacity. }
 \label{profile}
\end{figure}


Fig.1 shows schematically the conditions of our experiments. The experiments are made at zero pressure and at temperatures down to ~ 0.11 mK. The superfluid $^3$He-B energy gap at zero pressure is about 2 mK. Thus the quasiparticles gas is very dilute and consequently its heat capacity is very small. Near the wall the energy gap falls to zero. This region allows the existence of Majorana quasiparticles with zero energy gap. These quasiparticles can move in the 2D space along the surface. The heat capacity of bulk Bogolyubov quasiparticles and Majorana quasiparticles can be estimated as followings:

The Bogolyubov quasiparticles heat capacity in superfluid $^3$He-B  falls with temperature as \cite{VW,BB}:
\begin{equation}
  C_{bulk} \sim V \, P_F^2\,  (\frac{\Delta}{kT})^{3/2} \exp \;( - \frac{\Delta}{kT} )
   \label{Cbalk}
\end{equation}
where $P_F$ is the fermi momentum, $\Delta$ the superfluid gap $ \simeq 2kT_c$ and $V$ is the volume of the sample.

Owing to the zero gap the Majorana heat capacity follows the power law \cite{2b,Volest}:
\begin{equation}
  C_{maj} \sim A \, \xi \, P_F^2\,  (\frac{\Delta}{kT})^{-2},
   \label{Cmaj}
\end{equation}
where $\xi$ is the $^3$He-B coherence length and $A$ the surface
area of the sample.

The ratio of these heat capacities, including the numerical factors,
reads:
\begin{equation}
\frac{C_{maj}}{C_{bulk}} = \frac{\pi^3}{8\sqrt2} \frac{ \xi}{\lambda} \,
(\frac{\Delta}{kT})^{-7/2}\exp \, ( \frac{\Delta}{kT} )= F \, \frac{\xi }{\lambda} ,
   \label{Cratio}
\end{equation}
 where $\lambda $ is a geometrical parameter characteristic of the experimental cell, the ratio between the volume and the surface area.
The function $F$, introduced by this equation, is very useful since it indicates the temperature where the Majorana heat capacity begins to become significant. In a current experiments the parameter $\lambda/\xi $ is about $10^{-4}$. The crossover when the heat capacities of the Majorana and Bogolyubov quasiparticles are equal should occur when $ F= 10^4$, that is about 0.105 mK \cite{BunkovMaj}, which corresponds well with our experimental results.

We have used a  bolometer which consists of a closed copper box with a small orifice \cite{BunkovMaj}. The box has the form of a cylinder of 6 mm diameter and 5 mm height. The diameter of the orifice is about 0.2 mm.  The top and bottom plates of the cell are made from copper foils and the cylinder is turned copper. The volume is about 0.13 cm$^3$.  The bolometer is situated inside the chamber of a nuclear demagnetization refrigerator filled with superfluid  $^3$He at extremely low temperatures.  The only thermal reservoir in the bulk superfluid $^3$He consists of the thermal gas of Bogolyubov quasiparticles. The temperature inside the bolometer is determined by the balance of heat leak in  and cooling by the flow of quasiparticles flowing out of the orifice.  Because of the poor Kapitza resistance, the thermal conduction between the $^3$He and the walls of the bolometer can be taken as zero.

To calculate the heat capacity we measure the thermal response of the liquid in the bolometer to a calibrated heat pulse. To measure the temperature (density of Bogolyubov quasiparticles) we have used the vibrating wire resonator (VWR) techniques first described in \cite{lanc1}. The broadening of the VWR resonance is determined by the dissipative interaction with the Bogolyubov quasiparticles. (For a description of the VWR see Methods.)  There are two different temperature scales for ther frequency width of a VWR as a function of temperature, the Lancaster scale \cite{lanc1} and the Grenoble scale \cite{Grenterm}. Here we have used the Lancaster scale which fits our experimental data better.  After an energy deposition the quasiparticle density rises rapidly and then decreases with a time scale, $\tau_b$, of a few seconds, as shown in Fig. 2. The time constant $\tau_b$ is determined by the ratio of the volume of the bolometer and the arera of the orifice. If we take into account only the Bogolyubov quasiparticles, then $\tau_b$ changes only slowly with cooling since the heat capacity and energy flow through the orifice are both proportional to the quasiparticle density. However,  the time constant changes as soon as the additional heat capacity arising from the Majorana quasiparticles appears.  By observing the temperature-dependence of $\tau_b$ we have made the first confirmation of the existence of Majorana quasiparticles in superfluid $^3$He \cite{BunkovMaj,BunGaz}.

The quasiparticle density and  temperature  increase almost instantaneously after a heating pulse. The measured VWR broadening increases with some time delay due to the quality factor of the VWR.
The response time $\tau _{R}$ is inversely proportional to the baseline width of the VWR  ($W_{0}$);  $\tau_{R}=1/\left ( \pi W_{0} \right )$. Consequently, the time dependence of measured VWR response on the heating pulse at $t_{0}$ can be written \cite{DMB2}:
\begin{equation}
W\left ( t \right )=W_{0}+A\frac{\tau _{b}}{\tau _{b}-\tau _{R}}\left  [ \left(\exp (-\frac{t-t_{0}}{\tau _{b}} \right )-\exp \left (-\frac{t-t_{0}}{\tau _{R}} \right ) \right ]
	\label{W_mes}
\end{equation}
where  $A$ is the fitting parameter, the broadening of VWR immediately after the heat pulse, for the conditions of zero time delay of VWR thermometer.  We have fitted our data using this function as shown in Fig. 2 and tabulated the values of $A$. We note that the treatment of the data  in this letter is different from that in \cite{DMB2}. Here we have calculated the value of A separately for each event.

 \begin{figure}[htt]
 \includegraphics[width=0.5\textwidth]{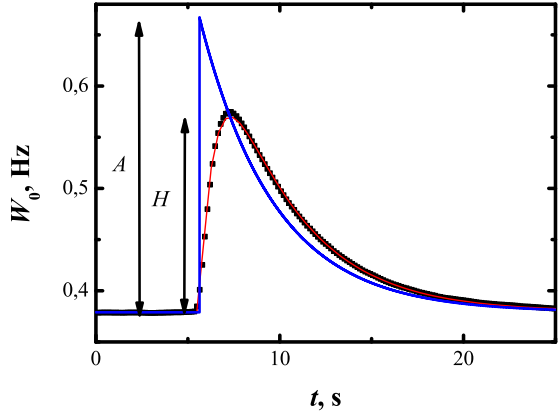}
 \caption{The evolution of the VWR width after a heating pulse. The red line is a fit of the experimental points by Eq. (8). The blue line is the corresponding quasiparticle density obtained by Eq. (7). H is the actual measured peak width increase and A the corresponding value assuming instantaneous VWR response.}
 \label{profile}
\end{figure}

 \begin{figure}[htt]
 \includegraphics[width=0.5\textwidth]{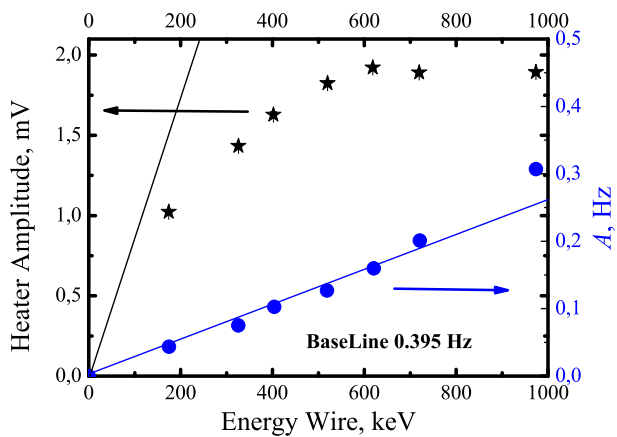}
 \caption{The amplitude of heater VWR oscillation just after the RF pulse (stars). The straight line shows the linear regime of  quasiparticle excitations, as discussed in the text.  Bullets shows the additional broadening of the thermometry VWR A  after heating pulses of different energies. The baseline width is about 0.39 Hz.}
 \label{profile}
\end{figure}

 \begin{figure}[htt]
 \includegraphics[width=0.5\textwidth]{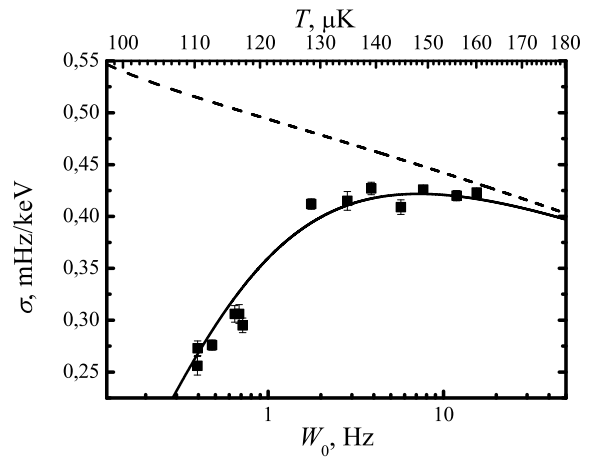}
 \caption{The calibration factor for a range of temperatures. The dashed line represents the variation expected from the Bogolyubov quasiparticles alone.   The solid line takes into account the additional heat capacity of the Majorana quasiparticles.}
 \label{profile}
\end{figure}

 \begin{figure}[htt]
 \includegraphics[width=0.5\textwidth]{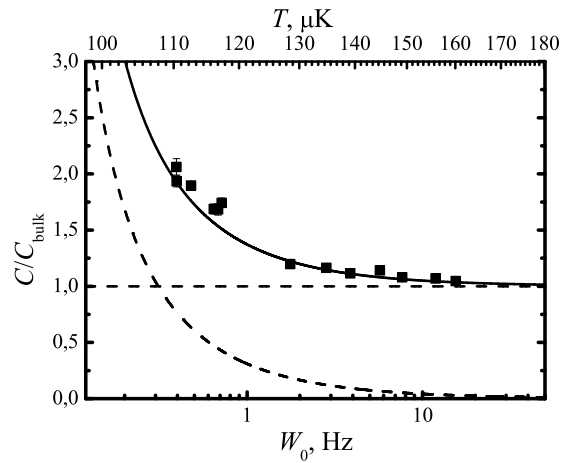}
 \caption{The  heat capacity ratio $C/C_{bulk}$ calculated from data shown in Fig.4. The line equal to 1 correspond to Bogoluybov QPs. The curved dashed line is that for the Majorana quasiparticles alone, and the solid line corresponds to a heat capacity of a both components.}
 \label{profile}
\end{figure}

Direct heating of superfluid $^3$He by an immersed heater is impossible owing to the very large Kapitza resistance between heater and liquid. Heat can be injected directly into the superfluid by exciting a VWR beyond the critical velocity for pair breaking as proposed in \cite{Heat}.   One should keep in mind that the electrically injected power is firstly transformed into mechanical energy of the VWR before being transferred to the liquid. By pair breaking the VWR motion creates the quasiparticles directly in the superfluid $^3$He each with a momentum of $\sim P_F$.  By integrating the voltage and current through the VWR we can calculate the deposited energy. The question is: how much of this energy is transformed to quasiparticles?  In our previous Dark Matter detector investigations \cite{DMB2} we surprisingly found a decrease of the sensitivity on cooling below 200 $\mu$K.  To explain this feature we suggested that part of VWR heater energy is lost through the intrinsic damping of the VWR. We suggested that a fraction of the energy $ W_{int}/W_{0} $ was lost due to VWR heating. Indeed, the parameter $ W_{int}$ was a fitting parameter. In our current experiments we have found that this mechanism cannot explain the increasing of time constant $\tau _{b}$, we observed earlier \cite{BunkovMaj}. Furthermore, we have analyzed the oscillation amplitude of the heating VWR just after the heating pulse. We find a clear saturation of the amplitude. In other words, the VWR reaches the critical velocity for cooper pair breaking as shown in Fig. 3.
Thus the mechanical energy transfer to the quasiparticles is very effective and we should assign all the dissipated VWR energy as contributing to the calibration pulses. The decrease of sensitivity, observed earlier, cannot be explained by mechanical loses, but rather by the additional heat capacity arising from Majorana particles. In Fig. 3  the calibration of the additional VWR broadening after the heating VWR excitation over a range of energies is also shown. By these measurements at different base temperature we are able to calculate the calibration factor of the bolometer, as shown in Fig. 4. The dashed line shows the calibration factor as a function of temperature (VWR damping) for the Bogolyubov quasiparticle heat capacity alone. However, we  see a significant deviation of the experimental data from this line, but if we also add the heat capacity of the Majorana quasiparticle (Eq. 2) we find good agreement between the experimental data and the theoretical curve.  In Fig. 5 the ratio of  the experimental data and the Bogolyubov quasiparticle heat capacity is shown. We know the volume of our bolometer, but the  surface of the walls has to be estimated taking into account the surface roughness. The walls of the bolometer are of turned copper (cylindrical body) and copper foils (ends). In both cases the roughness of the walls may contribute an effective area 5-10 times larger than the geometrical surface area. Electron microscope analysis of the surface also shows high roughness. We have used the surface area of the walls as a fitting parameter and found that it to be about 10 time bigger than the geometrical surface, in a reasonable agreement with the apparent roughness.

We should comment on the influence on our results of a possible layer of solid  $^3$He on the surfaces of the bolometer.  In our case the surface was covered by $^4$He during the condensation of the liquid $^3$He into the cell. Owing to the small surface-to-volume ratio it is enough to use the 0.1\% of $^4$He to completely prevent any solid $^3$He layer to form inside the bolometer. It was shown in  \cite{solid} that in the presence of fraction of solid $^3$He one should observed the two bolometer time constants arising from the the imperfect thermal contact between the solid and liquid $^3$He. We have not seen any such signature of solid $^3$He in our current experiments.

We thus can draw the conclusion that the existence of Majorana fermions is confirmed in this system by the very direct method of  measuring the added contribution of the Majorana heat capacity.

{\bf METHODS}

The VWR comprises a superconducting NbTi wire filament bent into an approximately semi-circular shape of a few mm diameter, with both ends firmly fixed \cite{Term}. The VWR is situated in a small magnetic field and driven by the Lorentz force generated by an AC current passed around the loop at close to the mechanical resonance frequency,  oscillating perpendicularly to its main plane with an rms velocity v. The motion is damped by frictional forces of total amplitude $F(v)$ arising mainly from momentum transfer to the quasiparticles of the surrounding superfluid with a magnitude proportional to the quasiparticle density. The quasiparticles density depends exponentially on the temperature so that the temperature dependence of the VWR damping can be written as \cite{lanc1}:
\begin{equation}
W\left ( T \right )=\alpha \exp \left ( -\frac{\Delta }{kT} \right )
	\label{W_exp}
\end{equation}
where $\alpha$ is constant determined by the microscopic properties of the liquid. The rapid exponential function ensures that  the VWR provides a very sensitive thermometer. Since it has been found that the damping depends slightly on the wire velocity  \cite{lanc2,DMB2}, we have used very small amplitudes of VWR excitation, which we have adjusted with temperature to ensure a constant amplitude  of the VWR resonator.

{\bf Acknowledgements}
We would like to thank G. Pickett,  G.E. Volovik and C. Winkelmann for very useful discussions. The authors are grateful to the ''Agence Nationale de la Recherche'' (France) for providing financial support within the MajoranaPRO project(ANR-13-BS04-0009-01). We are also grateful for the collaboration with Kazan Federal University under the auspices of the Russian Government Program for the Competitive Growth of Kazan Federal University.

{\bf Author contributions} All authors contributed equally to this work.



\begin{thebibliography}{99}

\bibitem{Majorana} M. Buchanan, Nature Phys., {\bf11}, 206, (2015).

\bibitem{DM} Chiu Man Ho, Robert J. Scherrer,
 Physics Letters B, {\bf 722}, 341 (2013).


\bibitem{Wil} F. Wilczek, Nature Phys., {\bf 5}, 614, (2009).

\bibitem{Volovik14}  M. Silaev and G.E. Volovik, JETP, {\bf 119}, 1042 (2014); arXiv:1405.1007 (2014).

\bibitem{VolovikUn}
 G. E. Volovik, "The Universe in a Helium Droplet", OXFORD Science Pbl., (2009)

\bibitem{graphin} M.I. Katsnelson and G.E. Volovik, arXiv:1310.3581 (2013).




\bibitem{1} V. Mourik { \it et al.}, Science {\bf 336}, 1003-1007 (2012).

\bibitem{11} S. Nadj-Perge { \it et al.}, Science {\bf 346}, 602-607 (2014).

\bibitem{2} S. Murakawa { \it et al.}, J. Phys. Soc. Jpn. {\bf 80}, 013602
(2011).

\bibitem{2b} T. Mizushima { \it et al.}, J. Phys.: Cond. Matter. {\bf 27}, 013203
(2015).

\bibitem{3} Y. Okuda and R. Nomura, J. Phys.:Cond Matter., {\bf 24}, 343201
(2012).



\bibitem{Volovik09}
 G. E. Volovik, JETP Lett. {\bf 90} 398 and 587 (2009).

\bibitem{Volovik10} G.E. Volovik,
JETP Lett. {\bf 91}, 201–205 (2010).


\bibitem{Andreev} A. F. Andreev, Sov. Phys, JETP,{\bf 19}, 1228
(1964).


\bibitem{suggest} T. Mizushima, K. Machida, J. Low Temp. Phys., {\bf 162}, 204 (2011).

\bibitem{Haplperin} C. Choi { \it et al.}, Phys. Rev. Lett., {\bf 96}, 125301 (2006).

\bibitem{VW} D. Vollhardt and P. Wolfle, The superfluid phases of helium 3, Taylor andFrancis, London (1990)

\bibitem{BB} C. Bauerle, Yu.M. Bunkov, S.N. Fisher, H. Godfrin,
Phys.Rev. B, {\bf 57}, 14381  (1998).

\bibitem{Volest} G. E Volovik, Privat communications, (2012).

\bibitem{BunkovMaj} Yu. M. Bunkov,
J. Low Temp. Phys., {\bf 175}, 385-394 (2014).

\bibitem{lanc1} S. N. Fisher, A. M. Guenault, C. J. Kennedy, and G. R. Pickett,
Phys. Rev. Lett. {\bf 63}, 2566 (1989).

\bibitem{Grenterm} C. B. Winkelmann, E. Collin, Yu. M. Bunkov, H. Godfrin,
J. of Low Temp. Phys., {\bf 135}, 3 , (2004).


\bibitem{BunGaz}  Yu. M. Bunkov and R. R. Gazizulin,
Appl. Mag. Res. {\bf 45}, 1219–1224 (2014).

\bibitem{DMB2} C. B. Winkelmann, J. Elbs,  Yu. M. Bunkov, E. Collin, H. Godfrin and  M. Krusius
Nucl. Instrum. Meth. {\bf A574}, 264-271 (2007).

\bibitem{Heat} D.I. Bradley, Yu.M. Bunkov, D.J. Cousins, M.P. Enrico,
S.N. Fisher, M.R. Follows, A.M. Guenault, W.M. Hayes,
G.R. Pickett, T. Sloan, Phys. Rev. Lett. {\bf 75}, 1887 (1995).



\bibitem{Term} C. Winkelmann, E. Collin, Yu. M. Bunkov, H. Godfrin
J. Low Temp. Phys.. {\bf 135}, 3, (2004)



\bibitem{solid}  J. Elbs, C. Winkelmann, Yu.M. Bunkov, E. Collin, H. Godfrin
J.  Low Temp. Phys.{\bf 148}, 749-753 (2007)


\bibitem{lanc2}S. N. Fisher, G. R. Pickett, R. J. Watts-Tobin,
J. Low Temp. Phys., {\bf 83}, 225 (1991).


 \end{thebibliography}
\end{document}